\documentstyle[prl,aps,twocolumn,epsf]{revtex}
\begin{document}
\draft \twocolumn[\hsize\textwidth\columnwidth\hsize\csname
@twocolumnfalse\endcsname
\title{Mixed phases in $U(N)$ superconductivity}
\author{M.A.Moore, T.J.Newman\cite{na}, A.J.Bray and S-K.Chin} 
\address{Theory Group, Department of Physics, 
 University of Manchester,\\ Manchester, M13 9PL, UK}
\maketitle
\begin{abstract}
  We consider a general class of type-II superconductors which are
  described by $N$ complex order parameters (OP), with an overall
  $U(N)$ symmetry for the energy functional.  In the lowest energy
  state, each OP adopts its own particular lattice structure, making
  the spatial variation of the overall condensate very complex.  We
  present detailed results for the case $N=2$. We also treat the limit
  $N \rightarrow \infty$, which allows us to resolve a recent
  controversy, by showing that all previous studies are 
  fundamentally flawed.
\end{abstract}
 ]

\narrowtext

\vspace{4 mm}

\section{Introduction}

It has been appreciated for many years that enlarging the symmetry
group of a given model can yield both new physical insights, and the
possibility of exact calculations. A familiar example is the
generalization of the Heisenberg model of ferromagnetism to the $O(N)$
model\cite{ma}, which allows an exact analysis of the critical
properties for $N \rightarrow \infty$. The large-$N$ extension of the
non-linear $\sigma$-model\cite{sig} has been similarly
fruitful\cite{signl}. The idea has also been applied recently to
non-equilibrium problems in turbulence\cite{wei}, phase
ordering\cite{po} and interface roughening\cite{ir}. The implicit
assumption involved is that the gross physical features are
insensitive to the size of the symmetry group -- an assumption which
is largely borne out in the above applications.

The first main point of this article is to emphasize that there is a
wide class of models whose physics does not change smoothly on
enlarging the symmetry group -- in the language of critical phenomena,
these are systems whose ordered phase is {\it spatially
  inhomogeneous}. We shall concentrate exclusively on one such system;
namely, the type-II superconductor, which condenses into the so-called
mixed-state -- a periodic array of flux lines\cite{ab}. However,
non-trivial effects will be apparent in the other systems within this
class (which may be fairly well characterized by having a wavelength
selection mechanism in their ordered phase.)

It is our intention to highlight the main physical changes which occur
when one generalizes the Landau-Ginzburg theory of superconductors to
a $U(N)$ theory involving $N$ complex order parameters (OP). The model
is not new -- in fact it has been studied many times in the past two
decades or so\cite{hlm,bnt,ba,lr,c1,c2,mn} -- however, the physics we
shall discuss here appears to have been overlooked in previous
studies\cite{ba,lr,c1,c2}. Although many of our conclusions apply to
arbitrary $N$, such is the richness of these systems, we shall discuss
in detail only the cases $N=2$ and $N \rightarrow \infty$.

Our use of the $U(N)$ symmetry is for convenience only, as it is the
smoothest continuation of the model away from $N=1$. Clearly for
higher values of $N$ there are an increasingly large number of
possible model Hamiltonians which one may construct from the $N$ OP's.
Within the field of unconventional superconductivity (as evidenced in
the heavy-Fermion system UPt$_{3}$), the commonly adopted model
is an $N=2$ Ginzburg-Landau theory with a reduced symmetry\cite{rev}.
The richness of the phase diagram is well-documented, with the
Abrikosov lattice giving way to exotic structures such as fractionally
quantized vortices, and flux textures. An application for $N=3$ can be
found in rotating He$_{3}$. Near to the normal-A1 transition,
the (18 component) OP may be reduced to three coupled complex
fields\cite{he3}, with the external angular velocity playing the role
of an applied magnetic field. These examples serve to show that the
strategy of enlarging the symmetry group of systems whose condensate
is spatially inhomogeneous is a `double-edged sword'. The advantage is
that one finds an increasingly rich variety of phases as one increases
$N$. The disadvantage is that one thereby loses immediate physical 
contact with the original model of interest ($N=1$).

The remainder of this article is dedicated to exhibiting the
strengths and weaknesses of the $U(N)$ model, in terms of its relation to
conventional superconductivity. In section II, we define the model
via the Landau-Ginzburg energy functional, and derive the mean field
theory for arbitrary $N$. In section III we study in detail the mean-field
solution for $N=2$ and demonstrate that the Abrikosov state is
unstable, giving way to a state of interlocked centered rectangular lattices.
In section IV we review the large-$N$ limit for this model. All previous
studies\cite{ba,lr,c1,c2} have assumed an Abrikosov state for one
condensed mode, and treated the remaining $(N-1)$ modes as massless.
We show that this state is unstable, and that the true ordered state
corresponds to a complicated structure in which each OP condenses into
a periodic state, such that the overall condensate density is spatially
constant. This leads to an identification of the transition with that
of the $O(2N)$ model of ferromagnetism (in the large-$N$ limit) in two fewer
dimensions. An immediate consequence of this is that the transition
is continuous and the lower critical
dimension of the system is $d_{l}=4$. We also discuss the subtleties of 
commuting the limit $N \rightarrow \infty$ with the thermodynamic limit.
It is of note that the large-$N$ limit may be solved with no 
approximations -- we adopt neither the London limit (gauge field
fluctuations only) nor the lowest Landau level (LLL) approximation 
(OP fluctuations only). We end with our conclusions in section V.

\section{Formulation of the model and mean field theory}

We define the model via the Landau-Ginzburg energy
functional\cite{fh}, which has the generic form (setting
$\hbar = c = 1$)
\begin{eqnarray}
\label{lg}
\nonumber {\cal H}[\psi_{i} & , & {\bf A}] = \int d^{3}x \Bigl \lbrace
\sum_{i}[(1/2m^{*})|{\bf D}\psi_i|^{2} + \alpha|\psi_i|^{2}]\\  
+ ( & \beta & /2)\sum_{i,j}|\psi_i|^{2}|\psi_j|^{2} + 
(1/2\mu_{0})({\bf H}-\nabla \times {\bf A})^{2}
\Bigr \rbrace \ ,
\end{eqnarray}
where ${\bf D}=-i\nabla - e^{*}{\bf A}$, $\lbrace \psi_{i}
\rbrace$ are a set of $N$ complex order parameters, ${\bf A}$ is the
vector potential, ${\bf H}$ is the external magnetic
field, $e^{*}$ is the effective charge, and $m^{*}$, $\alpha$ and
$\beta $ are phenomenological constants. It is useful to 
introduce dimensionless units\cite{fh}, such that the energy functional
(in the ordered phase) takes the form
\begin{eqnarray}
\label{lgdless}
\nonumber {\cal H}[\psi_{i},{\bf A}] & = & \int d^{3}x \Bigl \lbrace
\sum_{i}[|{\bf D}\psi_i|^{2} - |\psi_i|^{2}]\\ & & +(1/2)\sum
_{i,j}|\psi_i|^{2}|\psi_j|^{2} + ({\bf H}-\nabla \times {\bf A})^{2}
\Bigr \rbrace \ ,
\end{eqnarray}
where ${\bf D}=(1/i\kappa) \nabla - {\bf A}$.
The parameter $\kappa>1/\surd 2$ 
is the ratio of the London penetration
depth to the coherence length. In these units the critical external
field $H_{c2} = \kappa $. When generalizing the above expressions
to $d$ dimensions, one should
imagine the external field directed along $(d-2)$ dimensions, such that
it is transverse to the $(x,y)$ plane in which an Abrikosov-like
lattice structure may form.

The simplest analysis one can make of this system is mean field theory
which amounts to approximating the free energy by the saddle point
value of ${\cal H}$. This may be obtained by solving explicitly for
the classical fields which are solutions of the Landau-Ginzburg
equations
\begin{eqnarray}
\label{lge1}
{\bf D}^{2}\psi_i & = & \Bigl ( 1- \sum _{j}|\psi_j|^{2} \Bigr
)\psi_{i} \\ 
\label{lge2}
\nabla \times {\bf b} & = & (1/2)\Bigl ( \psi^{*} {\bf D} \psi +
\psi {\bf D}^{*} \psi ^{*} \Bigr ) \ ,
\end{eqnarray}
where ${\bf b}={\bf H}-\nabla \times {\bf A}$ is the microscopic magnetic
field.

Although it is not known how to solve these equations in general, an
exact analysis is possible very close to the critical field $H_{c2}$.
We shall briefly outline the solution in the case of $N=1$, and then
generalize the solution to arbitrary $N$. Further details may be
found in refs.\cite{ab,fh}. Near to $H_{c2}$ the OP is small, so that 
the first equation may be linearized, and at this order the gauge field
is just given by ${\bf A_{0}} = H_{c2}(0,x,{\bf 0}_{\perp})$. Thus, the OP may be
expressed in terms of Landau levels\cite{llqm}. In fact, the value of 
$H_{c2}$ itself is determined by associating criticality with the
LLL eigenvalue. For our purposes, the key point is that the first equation
may be rearranged in the form
\begin{equation}
\label{oeq}
D_{+}D_{-} \psi = 0 \ ,
\end{equation} 
where 
\begin{equation}
\label{dpm}
D_{\pm} = {\bf D}_{x} \mp i{\bf D}_{y} \ ,
\end{equation}
and we have used the fact that the OP components are constant in the
$(d-2)$ other directions.
Thus the ground state `wavefunctions' are characterized by
the identity 
\begin{equation}
\label{a0fi}
D_{-}\psi = 0 \ .
\end{equation}
These functions are 
the LLL, and have a degeneracy proportional to the area of the
system (transverse to the applied magnetic field). One is free
to construct different sets of functions from the LLL's. The most
elegant is due to Eilenberger\cite{ge}. The Eilenberger basis
functions span the space of the LLL, yet each basis function
is a doubly periodic function (essentially a particular
Abrikosov lattice). 

Given the above properties of the OP near the upper critical field,
it is possible to manipulate the second mean field equation
such that the r.h.s. is the curl of a vector. In this way, one
may integrate the equation to obtain Abrikosov's first fundamental
identity
\begin{equation}
\label{a1fi}
b({\bf r}) = H - (1/2\kappa) |\psi ({\bf r})|^{2} \ .
\end{equation}
We see that the magnetic field is reduced within the sample
by an amount proportional to the condensate density.

The scale of the condensate is set by Abrikosov's second identity.
This is a little more difficult to prove\cite{fh}, but is essentially
obtained by spatially averaging the first equation and calculating 
the first non-trivial corrections to eq.(\ref{a0fi}). The result is
\begin{equation}
\label{a2fi}
(1/\kappa)(\kappa - H)\langle |\psi |^{2} \rangle
+ ((1/2\kappa ^{2})-1)\langle |\psi |^{4} \rangle = 0 \ , 
\end{equation}
where $\langle ... \rangle $ denotes a spatial average.

Substituting this last relation into eq.(\ref{a1fi}) and spatially
averaging, we find
\begin{equation}
\label{magf}
B = \langle b \rangle = H - {(\kappa - H)\over (2\kappa ^{2}-1)\beta _{A} } \ ,
\end{equation}
where the Abrikosov ratio is given by 
\begin{equation}
\label{abrat}
\beta _{A} = \langle |\psi |^{4} \rangle / \langle |\psi |^{2} 
\rangle ^{2} \ge 1 \ .
\end{equation} 
One can translate this result in terms of the free energy, which one
finds to be proportional to $-1/\beta _{A}$, indicating that
the system condenses into a state which minimizes $\beta _{A}$ under
the constraint that it is an Eilenberger function. This state turns
out to be the well-known triangular lattice, for which 
$\beta _{A} \simeq 1.1596$.

Now we come to the general $U(N)$ case as described by eqs.(\ref{lge1})
and (\ref{lge2}). It is easy to show that most of the previous analysis
follows through in a trivial way. Each OP component must satisfy
$D_{-}\psi_{i} = 0$ (although it is important to remember that a given
component may be zero). The first Abrikosov identity is
generalized to 
\begin{equation}
\label{ga1fi}
b({\bf r}) = H - (1/2\kappa)\sum _{i} |\psi _{i}({\bf r})|^{2} \ .
\end{equation}
There also exist $N$ relations setting the relative scales of the OP
components. These have the form
\begin{equation}
\label{ga2fi}
(1/\kappa)(\kappa - H)\langle |\psi_{i}|^{2} \rangle
+ ((1/2\kappa ^{2})-1)\sum _{j}\langle |\psi_{i}|^{2} 
|\psi_{j}|^{2}\rangle = 0 \ . 
\end{equation}

This allows us to spatially average eq.(\ref{ga1fi}) to find the
relation between the magnetic flux density and the magnetic field:
\begin{equation}
\label{gmagf}
B = H - {(\kappa - H)\over (2\kappa ^{2}-1)\beta _{g}(N) } \ ,
\end{equation}
where the generalized Abrikosov ratio is given by 
\begin{equation}
\label{gabrat}
\beta _{g}(N) = { \sum_{i,j}\langle |\psi _{i}|^{2} |\psi _{j}|^{2} 
\rangle \over \left (\sum_{i} \langle |\psi_{i}|^{2} \rangle  
\right ) ^{2}} \ge 1 \ .
\end{equation} 

Again, one may show that the minimal free energy is to be found by
minimizing $\beta _{g}$, with the constraint that each condensed
OP component is either zero, or an Eilenberger function. For general
$N$ one soon appreciates that such a minimization is non-trivial as
the $N$ components jostle for favorable positions and lattice
structures within the `primitive cell'. For this reason we concentrate
in the next section on the simplest case.

\section{Analysis of the case $N=2$}

We are now only concerned with two OP's  $\psi _{1}$ and $\psi _{2}$.
By reducing the $U(2)$ symmetry we could make contact with the two-OP
models of UPt$_{3}$\cite{rev} for which a similar mean field analysis has been
performed\cite{mf2op}. Our purpose here is to exemplify the physics of
these multi-component systems -- this will be of great benefit in our
analysis of the large-$N$ limit in the following section.

A few words concerning the Eilenberger basis\cite{ge}
are required at this point.
As mentioned before, the Eilenberger functions $\phi ({\bf r}|{\bf r}_{0})$
satisfy $D_{-}\phi = 0$. The amplitude of $\phi$ is a doubly periodic
function, with a fundamental cell scaled so as to have unit length
in the $x$ direction, and a periodicity vector $(\zeta,\eta)$, where
$\eta $ is fixed by the condition of flux quantization. The label 
${\bf r}_{0}$ simply fixes the lattice position in space. The functions
are normalized such that $\langle |\phi|^{2} \rangle = 1$. 
The functions also have the symmetry property 
$\phi ({\bf r}|{\bf r}_{0}) = \exp[2\pi i(y_{0}/\eta)x]\phi({\bf r}+
{\bf r}_{0}|{\bf 0})$. This allows one to recast the normalization
condition as a completeness relation:
\begin{equation}
\label{com}
\int \limits _{\rm cell} d^{2}r_{0} |\phi ({\bf r}|{\bf r}_{0})|^{2} = \eta \ .
\end{equation}
It is useful to define the integrals
\begin{equation}
\label{ints}
I({\bf r}_{1},{\bf r}_{2}) = \int \limits _{\rm cell} d^{2}r 
|\phi ({\bf r}|{\bf r}_{1})|^{2}|\phi ({\bf r}|{\bf r}_{2})|^{2} \ . 
\end{equation}

Returning to our expression for $\beta _{g}$ in the case $N=2$, we have
to minimize the expression
\begin{equation}
\label{betag2}
\beta _{g}(2) = { \langle |\psi _{1}|^{4}\rangle + \langle |\psi _{2}|^{4} 
\rangle + 2 \langle |\psi _{1}|^{2}|\psi _{2}|^{2}\rangle 
\over \left ( \langle |\psi_{1}|^{2}\rangle + \langle 
|\psi_{2}|^{2} \rangle  \right ) ^{2}} \ .
\end{equation}
One may show that a minimum may only exist for either one of the OP's
being zero, or else for each OP to be equivalent (up to a relative spatial
shift). In the former case the value of $\beta _{g}$ will be the 
Abrikosov value ($\simeq 1.1596$). To investigate the latter case, we set
$\psi _{1} = A \phi ({\bf r}|{\bf 0})$ and  
$\psi _{2} = A \phi ({\bf r}|{\bf r}_{0})$ which reduces the
task of minimizing $\beta _{g}$ to that of finding the lattice
type and the relative shift ${\bf r}_{0}$ which minimize
\begin{equation}
\label{nbetag2}
\beta _{g}(2) = I({\bf 0},{\bf 0}) + I({\bf 0},{\bf r}_{0}) \ .
\end{equation}

We have numerically investigated the above expression, restricting
our search to lattices within the class of centered rectangular 
structures (which corresponds to choosing $\zeta=1/2$). This class
includes square and triangular lattices. Somewhat surprisingly the
minimal energy solution corresponds to each OP component adopting a 
lattice with a primitive cell with opening angle $\theta $ 
equal to $15^{\rm o}$.
Figures 1 and 2 show contour plots of the two OP components.
Regions of lighter shade correspond to higher values of the OP.
It is more illuminating to plot
the overall condensate $|\phi _{1}|^{2} + |\phi_{2}|^{2}$, as is shown
in Fig. 3. We then see
a surprisingly rich structure. Maxima of the condensate correspond
to regions of minimal magnetic flux and {\it vice versa}. The energy
for this arrangement corresponds to the value  $\beta _{g}(2) \simeq 1.0062$,
which is very close to the lower bound of unity. 
             
Our initial guess was that the OP components would adopt square
lattices ($\theta = 45^{\rm o}$), 
as this arrangement has a higher symmetry. We plot the
components and the overall condensate in Figs. 4, 5 and 6 for this
case. The energy of the square lattice arrangement is only 
fractionally higher with $\beta _{g}(2) \simeq 1.0075$. 

\newpage

\begin{figure}[tbp]
\centerline{\epsfxsize=5.0cm
\epsfbox{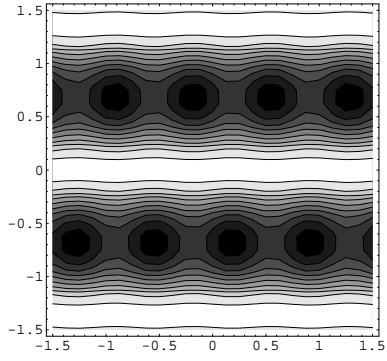}}
\vspace{0.5cm}
\caption{\label{min1} $|\phi_{1}|^{2}$ for minimum energy lattice.}
\end{figure}
             
\begin{figure}[tbp]
\centerline{\epsfxsize=5.0cm
\epsfbox{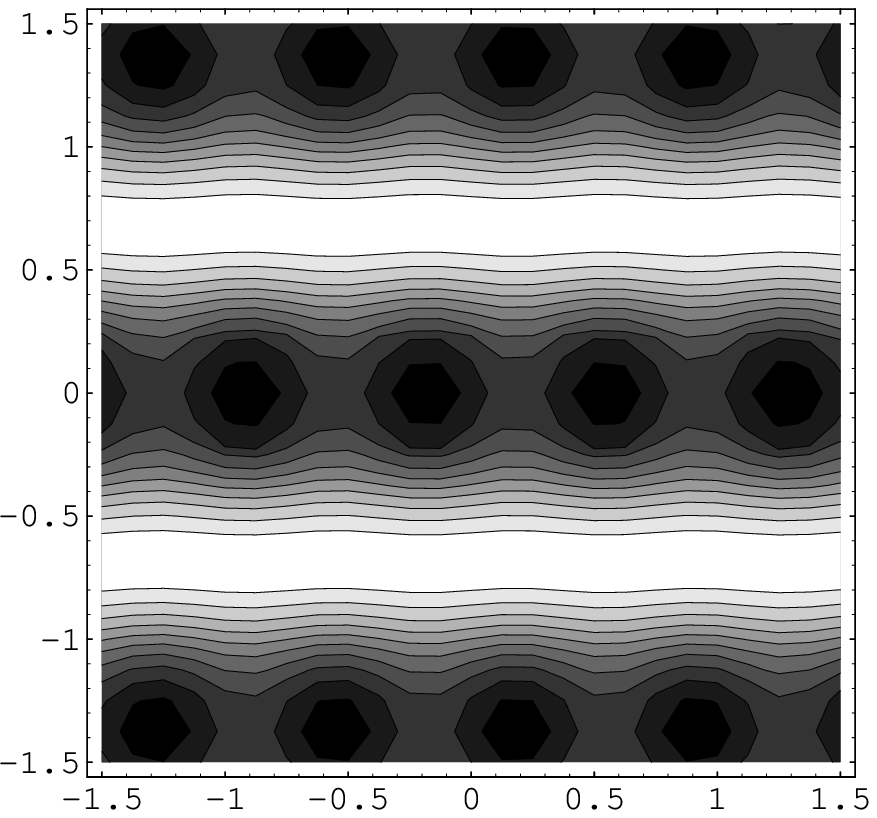}}
\vspace{0.5cm}
\caption{\label{min2} $|\phi_{2}|^{2}$ for minimum energy lattice.}
\end{figure}

\begin{figure}[tbp]
\centerline{\epsfxsize=7.5cm
\epsfbox{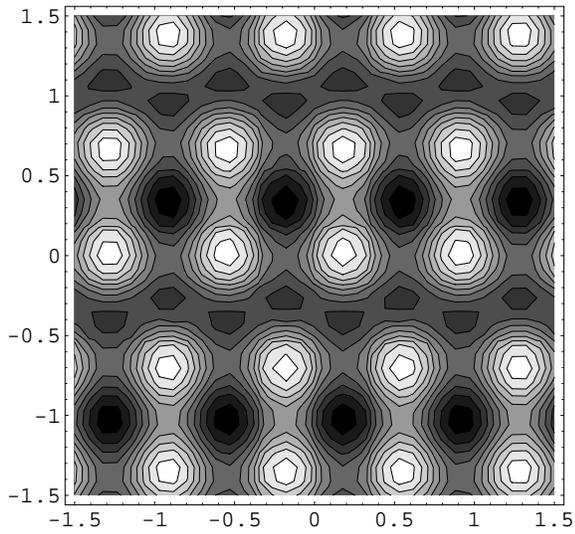}}
\vspace{0.75cm}
\caption{\label{min12}Condensate $|\phi_{1}|^{2}+|\phi_{2}|^{2}$ for
minimum energy lattice.}
\end{figure}

\begin{figure}[tbp]
\centerline{\epsfxsize=5.0cm
\epsfbox{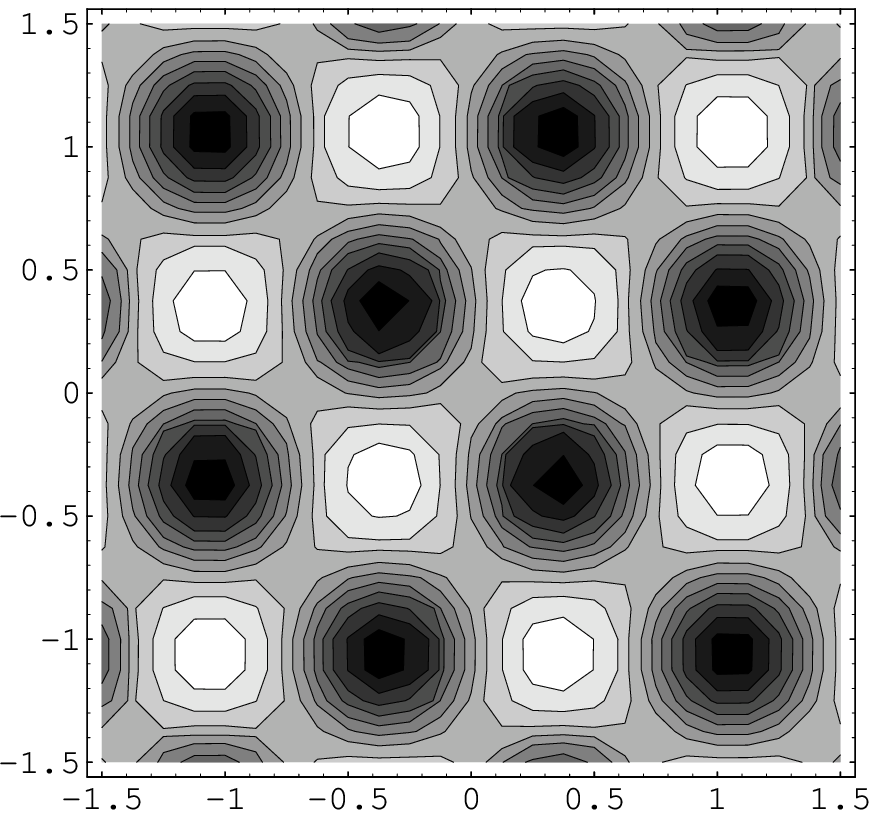}}
\vspace{0.5cm}
\caption{\label{sqr1} $|\phi_{1}|^{2}$ for square lattice.}
\end{figure}

\begin{figure}[tbp]
\centerline{\epsfxsize=5.0cm
\epsfbox{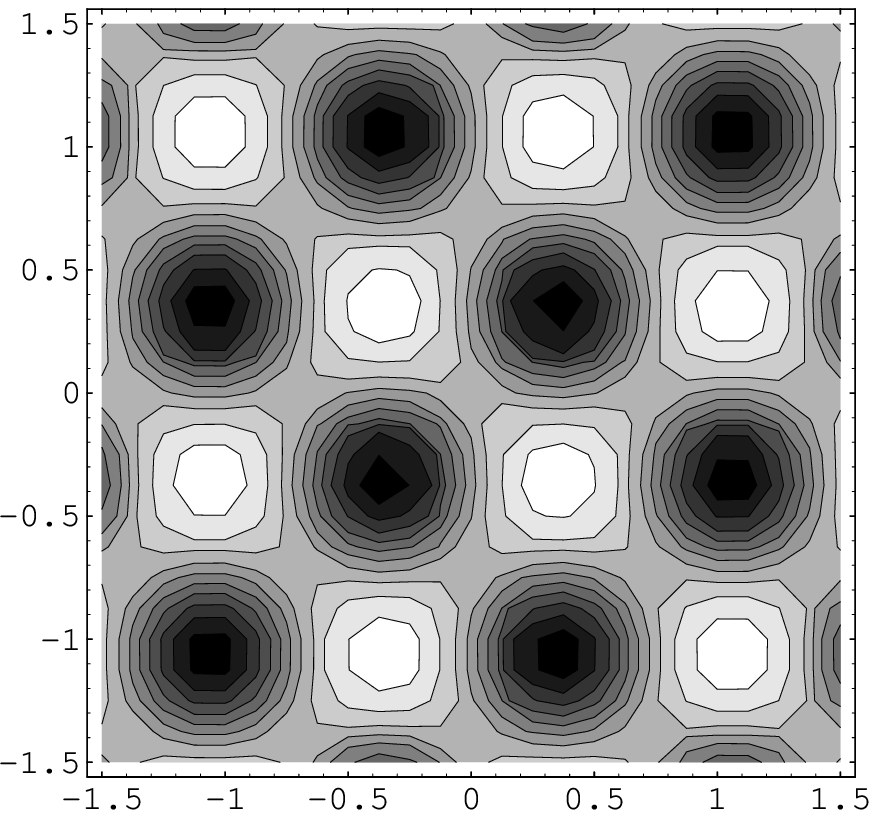}}
\vspace{0.5cm}
\caption{\label{sqr2} $|\phi_{2}|^{2}$ for square lattice.}
\end{figure}

\begin{figure}[tbp]
\centerline{\epsfxsize=7.5cm
\epsfbox{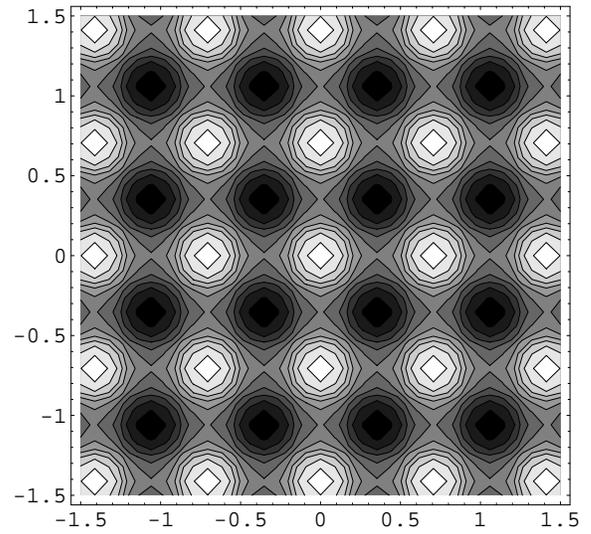}}
\vspace{0.75cm}
\caption{\label{sqr12}Condensate $|\phi_{1}|^{2}+|\phi_{2}|^{2}$ for
square lattice.}
\end{figure}

\newpage

In Fig.7 we show
the generalized Abrikosov number as a function of angle. Interestingly,
the triangular lattices correspond to maximal values of $\beta _{g}$.
In principle, there may be an even lower energy OP arrangement corresponding
to a lattice with an oblique primitive cell lying outside the
class of centered rectangular structures. This is a large space
within which to search for minima, and we have not pursued this 
possibility.

\begin{figure}[tbp]
\centerline{\epsfxsize=7.0cm
\epsfbox{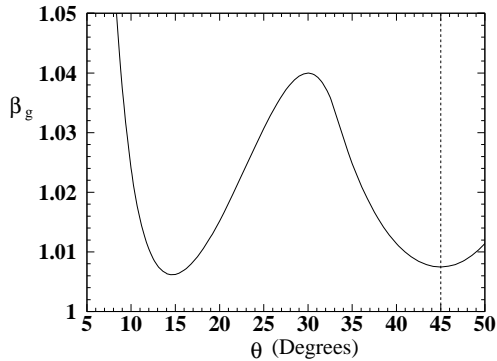}}
\vspace{0.7cm}
\caption{\label{beta}The function $\beta _{g}(2)$ plotted against $\theta $.
The function is symmetric about $45^{\rm o}$.}
\end{figure}

From this analysis, we see that in general
the $U(N)$ models will condense into complicated structures in which
each component adopts a particular lattice configuration, such that 
the overall condensate density is as smoothly varying as possible.
The calculation of these structures becomes increasingly difficult
as one increases $N$, and we shall desist from any explicit calculations
In the next section we turn to the most
important part of the article -- namely the explicit solution of the 
$U(N)$ model in the large-$N$ limit. 

\section{The large-$N$ limit}

As hinted at in the Introduction, one of the more compelling reasons
for generalizing models to higher symmetry groups is to allow
exact solutions in the large-$N$ limit. This has proven to be a very
useful tool in many contexts\cite{ma,signl,wei,po,ir}. In the present
case, the large-$N$ limit of the $U(N)$ model of type-II superconductors
(in an external field) has been examined by several 
authors\cite{ba,lr,c1,c2}. [One should draw a clear line between these
calculations, and those concentrating on the zero field case\cite{hlm}, 
which is of interest in liquid crystals and also as an exemplification
of a fluctuation induced first-order phase transition\cite{cw}.]
Originally studied in 1985\cite{ba}, it was found
that below six dimensions, the mean-field continuous transition of Abrikosov 
becomes first-order -- all the way down to a lower critical dimension of 
$d_{l}=2$. This calculation relied upon a proof by contradiction, as an
explicit solution of the model is extremely difficult.
An independent study\cite{lr} was made in 1995 using an
Ansatz to solve the model. It was found that   
the transition was continuous below six dimensions, and that $d_{l}=4$.
This result was challenged\cite{c1} on the grounds 
that the Ansatz was physically
unreasonable and that the original calculation\cite{ba} was definitely correct.
A spirited response\cite{c2} was made, defending the Ansatz, 
but admitting that there was no obvious flaw in the original proof 
by contradiction\cite{ba}. This is quite an extraordinary 
situation regarding a model whose {\it raison d'etre} is its
solvability! 

We shall resolve this state of confusion
using a very simple physical idea that has emerged from the previous
two sections. Each of the previous studies
follows the conventional route of integrating out $(N-1)$ of the OP components,
regarding them as Goldstone modes. This is precisely what one does in the
conventional $O(N)$ model of ferromagnetism for instance\cite{ma,amit}. 
The reason it
is done there is that the condensed OP is spatially homogeneous,
and it makes good sense to globally rotate the OP's such that the
condensate exists exclusively in one component (denoted as longitudinal),
and to treat the transverse components as a source of massless fluctuations.
Such a course of action is
inadmissible in the present model of $U(N)$ superconductivity. Each component
of the condensate is spatially inhomogeneous and no global rotation is
possible to make $(N-1)$ Goldstone modes. One has to treat all the modes
as potentially condensed. For finite $N$ this would lead to a totally
intractable problem. However, for $N \rightarrow \infty$ we can take
advantage of the completeness relation of Eilenberger functions to demonstrate
that although each OP component is a doubly periodic function,
the overall condensate is a constant. This drastically simplifies the
analysis, and one may show that the solution is self-consistent when one
takes into account both higher Landau levels, and gauge field
fluctuations. 

Before turning to the self-consistent large-$N$ limit, we shall obtain
the main idea by studying the limit of large $N$ at mean field 
level. The task is to minimize $\beta _{g}(N)$ as given in
eq.(\ref{gabrat}). We see that by choosing each component to lie
in a different Eilenberger state, the sums over components
tend (in the limit of large $N$) to integrals over the basis
label ${\bf r}_{0}$. We may then invoke the completeness relation
given in eq.(\ref{com}) which reduces these component sums to 
constants. Hence the spatial averages are trivial, and we see that
$\beta _{g}$ is saturated at its lower bound of unity. 

It is a remarkable fact that this particular solution actually
solves the mean field equations throughout the mixed phase, and not
just in the vicinity of the $H_{c2}$ line. One can see this
straightforwardly: since $\sum _{i} |\psi_{i}|^{2}$ is a constant,
the first mean-field equation is exactly solved by the Eilenberger
functions. The higher Landau levels always remain hard modes
as one decreases the temperature, so never contribute to the
condensed OP. At low enough temperatures, the Meissner transition
will occur, and the condensate 
$\sum _{i} |\psi_{i}|^{2}$ will saturate at the value
$2\kappa H$.

Two interesting points follow. In 
the previous section we found that even for two OP components, the
condensed state was very smooth (in terms of the overall condensate).
In that sense it is already very similar to the large-$N$ limit
in which the condensate is exactly constant. Thus we expect very
similar quantitative physics as one increases $N$ in the range
$[2,\infty]$. 
Note also the huge degeneracy of lattice structures
underlying this solution. Although each OP component is
restricted to be an Eilenberger function with a 
particular shape of unit cell, there is no energetic
selection of that unit cell for $N \rightarrow \infty$,
since the completeness relation is independent of this.
Presumably the ground-state OP configurations for large
but finite $N$ provide a means of determining the large-$N$
configurations, by smoothly continuing $N \rightarrow \infty$.

The self-consistent large-$N$ treatment is a non-trivial extension
of the above calculation, as it takes into account fluctuations
about mean-field theory (albeit in a rather crude fashion.) We shall
find that the main characteristics of the mean field state are stable to 
these fluctuations
for $d>d_{l}=4$. For $4<d<6$ the fluctuations give rise to $d$-dependent
exponents, which cross over to mean-field values above six dimensions.
In fact, from the structure of the theory, one sees that the results
are identical to those of the large-$N$ limit of the $O(2N)$ model
of ferromagnetism in two dimensions fewer. We can understand this as
the correlations of the OP's are frozen in the $(x,y)$ plane due
to each OP component having formed a lattice state, so only transverse
fluctuations (in the remaining $(d-2)$ dimensions) can become
critical.

[We note here that these results are similar to those of ref.\cite{lr},
but we must stress that the physical condensed state is completely
different in the two cases. In ref.\cite{lr} there is no explicit
transverse scale as only one OP component is condensed, and is
taken {\it ad hoc} to be constant (which is not even a solution
of the saddle-point equations\cite{c1}); whereas in our solution,
each OP component has condensed into a lattice state with its own
magnetic length scale built in. Real thermal fluctuations about these
two states will be of totally different natures.]

The large-$N$ limit is often derived diagrammatically\cite{ma,amit}.
When one can describe the physics in terms of a longitudinal
mode and $(N-1)$ transverse modes, this approach is particularly
transparent. However, in the present case we have $N$ condensed
modes, so it is better to use an alternative method; namely
to introduce an auxiliary field $\chi $ (via a Hubbard-Stratonovich
transformation) which will allow us to make the large-$N$ limit
explicit.

In the presence of fluctuations, we must take a step back from
eq.(\ref{lg}) and consider the partition function 
$Z = \int {\cal D}{\bf A} {\cal D}\psi_{i} 
\exp [-{\cal H}']$, where
\begin{equation}
\label{newen}
{\cal H}' = {\cal H} - \int d^{d}r \sum _{i} (J_{i}^{*}\psi _{i} + 
{\rm c.c}) \ .
\end{equation}
The source terms $J_{i}$ are added in order to derive the
equation of state in the ordered phase. It will turn out that
each $J_{i}$ is proportional to an Eilenberger function.

Introducing the field $\chi $ allows us to rewrite the 
partition function in the form
\begin{equation}
\label{npf}
Z[J] = \int {\cal D}{\bf A} {\cal D}\chi  {\cal D}\psi_{i} 
\exp [-{\cal H}''] \ ,
\end{equation}
where
\begin{eqnarray}
\label{neweren}
\nonumber & {\cal H}'' & = \int d^{d}r \bigl \lbrace N\chi ^{2}/2\beta
+ N({\bf H}-\nabla \times {\bf A})^{2} \\ 
& + & \sum_{i} \bigl [ (\alpha+i\chi) |\psi_i|^{2} + (1/2m^{*})|{\bf
  D}\psi_i|^{2} - J_{i}^{*}\psi _{i} - J_{i}\psi _{i}^{*} \bigr ]
\bigr \rbrace \ .
\end{eqnarray}
We have scaled the magnetic field and the vector potential 
so as to extract a clean factor of $N$ in the first two terms\cite{ba}.
(This entails the rescalings $e^{*} \rightarrow e^{*}/ \surd N$,
and $\beta \rightarrow \beta /N$ for consistency).

As we have already indicated, the self-consistent large-$N$ limit 
for this problem
constitutes a formidable analytic challenge. In fact, it is not possible
to solve the problem without resort to some external resource, whether
it be an Ansatz, or a piece of physical insight.
We shall utilize the latter, thanks to the lessons we have learned
both in the $N=2$ case, and also in the mean field analysis of the
large-$N$ limit. Just to reiterate, at mean
field level the OP components each condense into an Eilenberger state
such that the overall condensate is spatially homogeneous, and consequently
the magnetic flux is spatially homogeneous also.  To proceed we take the 
simplest possible line. Namely, that the self-consistent treatment retains
these features of spatial homogeneity, but that the fluctuations
renormalize the mass $\alpha $, resulting in a shift of $T_{c}$. Our task
is to show that this is a consistent solution of the problem. The 
alternative Ansatz is to condense only one OP component\cite{ba,lr,c1,c2}.
Rewriting the energy functional in terms of Eilenberger functions allows one
to prove that such a state is energetically unstable\cite{mn1}.

To select the physically motivated condensed state we must choose the source 
terms to force
each component into a (spatially shifted) Eilenberger function. Thus we
write $J_{i} = u\phi_{i} = u\phi ({\bf r}|{\bf r}_{i})$, where $u$ is
complex. The homogeneity of the magnetic field allows us to write
${\bf B}=B(0,0,{\hat {\bf r}}_{\perp})$. Also it is convenient to set
$t=r+i\chi $. The saddle point value of $\chi $ is purely imaginary, such that
the effective mass $t$ is purely real. The energy function now takes the 
explicit form
\begin{eqnarray}
\label{newesten}
\nonumber {\cal H}'' & = & \int d^{d}r \bigl \lbrace - N(t-r)^{2}/2\beta
+ N(H-B)^{2} \\ 
& + & \sum_{i} \bigl [ t|\psi_i|^{2} + (1/2m^{*})|{\bf
  D}\psi_i|^{2} - u\phi_{i}^{*}\psi _{i} - u\phi_{i}\psi _{i}^{*} \bigr ]
\bigr \rbrace \ ,
\end{eqnarray}
where now ${\bf A} = B(0,x,{\bf 0}_{\perp})$. 
Each individual OP component is decoupled, and may be associated with
a partition function $z(t,u) = \int {\cal D}\psi _{i} \exp(-f)$
where 
\begin{equation}
\label{indiv}
f[\psi _{i}] = \int d^{d}r \bigl [ \psi_{i}^{*}{\hat M}\psi_{i}
- u\phi_{i}^{*}\psi _{i} - u\phi_{i}\psi _{i}^{*} \bigr ] \ ,
\end{equation}
where ${\hat M} = -(1/2m^{*}){\bf D}^{2}+t$, which has eigenvalues 
\begin{equation}
\label{eigenval}
\lambda _{{\bf k},n} = {k^{2}\over 2m^{*}} + t + \left (n+{1\over 2}
\right ){e^{*}B\over m^{*}} \ ,
\end{equation}
where the momentum ${\bf k}$ exists in the $(d-2)$ dimensions transverse
to the $(x,y)$ plane, and $n \in [0,\infty]$ labels the Landau levels.

We now write the condensed part of the OP explicitly, along with its
fluctuation: $\psi _{i} = w\phi _{i} + {\tilde \psi}_{i}$. 
The prefactor $w$ is chosen to ensure that the fluctuation piece 
${\tilde \psi}_{i}$ has zero mean. Substituting this into eq.(\ref{indiv})
and performing the integrals over the Eilenberger functions, one finds
that the terms linear in ${\tilde \psi}_{i}$ vanish so long as
one chooses $w=u/(t+e^{*}B/2m^{*})$. In this case the energy functional for
each component becomes
\begin{equation}
\label{indiv1}
f[\psi _{i}] = -{V|u|^{2}\over 2(t+e^{*}B/2m^{*})}
+ \int d^{d}r \ {\tilde \psi}_{i}^{*}{\hat M}
{\tilde \psi}_{i} \ ,
\end{equation}
where $V$ is the volume of the system.

Returning now to the energy functional, we have from eqs.(\ref{newesten})
and (\ref{indiv1}), along with assumed spatial constancy of $\chi $ and
$B$,
\begin{eqnarray}
\label{energyfun}
\nonumber
{\cal H}'' = & - & NV \Biggl [ {(t-r)^{2}\over 2\beta} - (H-B)^{2} + 
{|u|^{2}\over 2(t+e^{*}B/2m^{*})}\Biggr ] \\
& + & \int d^{d}r \ {\tilde \psi}_{i}^{*}{\hat M}
{\tilde \psi}_{i} \ .
\end{eqnarray}

The integrals over the fluctuation fields $\lbrace {\tilde \psi}_i \rbrace$
are easily done, and we may re-exponentiate the resulting determinant
to give the final result 
\begin{eqnarray}
\label{energyfinal}
\nonumber
{\cal H}'' = & - & NV \Biggl [ {(t-r)^{2}\over 2\beta} - (H-B)^{2} + 
{|u|^{2}\over 2(t+e^{*}B/2m^{*})} \\
& - & (D_{L}/A) \int {d^{d-2}k\over (2\pi)^{d-2}}\sum _{n} \log 
(\lambda _{{\bf k},n})\Biggr ] \ ,
\end{eqnarray}
where $D_{L}$ is the Landau degeneracy of each level, which is equal
to $e^{*}BA/2\pi$, where $A$ is the area of the system in the $(x,y)$ plane.

Now that we have a clean factor of $N$ throughout, we may use steepest 
descents to determine the self-consistent values of the auxiliary 
variable $t$ , and also the magnetic field $B$. Differentiating ${\cal H}''$
with respect to $t$ yields
\begin{eqnarray}
\label{saddle}
\nonumber
{(t-r)\over \beta} & - & { |u|^{2} \over 2( t+e^{*}B/2m^{*})^{2}}\\ 
& & \ \ \ \ - {e^{*}B\over 2\pi}\int {d^{d-2}k\over (2\pi)^{d-2}}\sum _{n}
{1\over \lambda _{{\bf k},n}} = 0 \ .
\end{eqnarray}
It is convenient to define $\xi ^{-2} = t+e^{*}B/2m^{*}$, since $\xi $ can
be seen to play the role of the correlation length of the system. 
The relation between $w$ and $u$ then assumes the form $w=u\xi^{2}$.
One may then rewrite eq.(\ref{saddle}) as
\begin{eqnarray}
\label{self-conxi}
\nonumber
& \xi ^{-2} & = r + e^{*}B/2m^{*} + \beta |w|^{2}/2 \\ 
 + & \beta & {e^{*}B\over 2\pi}\int {d^{d-2}k\over (2\pi)^{d-2}}\sum _{n}
{1\over [ k^{2}/2m^{*} + ne^{*}B/m^{*} + \xi ^{-2} ]} \ .
\end{eqnarray}

This equation encapsulates most of the information about the phase transition.
Above the transition, we can set the condensate `amplitude' $w$ to zero
in (\ref{self-conxi}). The resulting equation defines the renormalized 
critical temperature $T_{c}$ through
the defining condition of criticality $\xi \rightarrow \infty$.
At this point the bare quantity $r=T-T_{c}^{0}$ is equal to $T_{c}-T_{c}^{0}$.
We therefore have the explicit shift as
\begin{eqnarray}
\label{tcshift}
\nonumber
T_{c} & = & T_{c}^{0} - e^{*}B/2m^{*} \\ 
 - & \beta & {e^{*}B\over 2\pi}\int {d^{d-2}k\over (2\pi)^{d-2}}\sum _{n}
{1\over [ k^{2}/2m^{*} + ne^{*}B/m^{*} ]} \ .
\end{eqnarray}
As usual, the fluctuations drive the critical temperature to a lower
value. The $T_{c}$ shift diverges for $d<4$, suggesting the identification
of the lower critical dimension as $d_{l}=4$.

On the low temperature side of the transition we can remove the source
field $u$, which leaves a non-zero condensate $w$ only if the correlation
length is infinite. We therefore have from (\ref{self-conxi}), an equation
for the condensate amplitude:
\begin{eqnarray}
\label{amplitude}
\nonumber
0 & = & r + e^{*}B/2m^{*} + \beta |w|^{2}/2 \\ 
 + & \beta & {e^{*}B\over 2\pi}\int {d^{d-2}k\over (2\pi)^{d-2}}\sum _{n}
{1\over [ k^{2}/2m^{*} + ne^{*}B/m^{*} ]} \ .
\end{eqnarray}
Comparing eqs.(\ref{tcshift}) and (\ref{amplitude}) we find the exact
relation $|w|^{2} = 2(T_{c}-T)/\beta$ which immediately yields the
OP exponent $\beta=1/2$, and self-consistently confirms the existence of
a continuous transition.  

We can also identify the correlation length exponent by examining
(\ref{self-conxi}) as $T \searrow T_{c}$. Eliminating the bare critical
temperature from (\ref{self-conxi}) using (\ref{tcshift}), and evaluating
the resulting integral for large $\xi $ leads to the expression
\begin{equation}
\label{corrlen}
c_{1}\xi ^{-2} + \beta e^{*}B c_{2} \xi ^{4-d} = T - T_{c} \ .
\end{equation}
The constant $c_{1}$ has a contribution from all Landau levels bar the
lowest. The fluctuation dominated term $\sim \xi ^{4-d}$ and arises 
solely from the LLL. As the correlation length diverges for
$T \searrow T_{c}$, we see that the first term on the left hand side
dominates for $d>6$, whereas the second term dominates for $4<d<6$. 
This leads to the result $\xi \sim (T-T_{c})^{-\nu}$, with $\nu = 1/2$
for $d>6$ (the mean field result), and $\nu = 1/(d-4)$ for $4<d<6$
(confirming $d_{l}=4$). 

These results are identical to those obtained for the $O(2N)$ model
of ferromagnetism, but in two fewer dimensions. This may be understood
from examination of the self-consistent relation for $\xi$ given in
(\ref{self-conxi}). Apart from the sum over Landau levels, this equation
is exactly that which would be obtained for a $(d-2)$-dimensional
$O(2N)$ model. The critical modes only exist in the $(d-2)$ 
dimensions transverse to the Landau levels. The modes in the 
$(x,y)$ plane contain the frozen length scale associated with the
formation of the underlying lattice structure of the OP's.

In the above expressions, we have left the value of the magnetic field
$B$ undetermined. However, this is given self-consistently by minimizing
the energy functional in (\ref{energyfinal}) with respect to $B$.
We shall not write the expression explicitly, but it is
noteworthy that the integral appearing from the fluctuations is
strongly divergent and must be regularized by introducing some
microscopic cut-off procedure (for $d>4$), such as adding higher
derivative terms not present in our original Landau-Ginzburg energy
functional.

As a final remark in this section, we should point out that these results
are insensitive to the order in which one takes the thermodynamic limit, 
and the limit $N \rightarrow \infty$. The above analysis has implicitly
assumed the thermodynamic limit, which is the correct `physical' choice.
However, had one taken the thermodynamic limit second, by fixing the number
of vortices and then taking $N \rightarrow \infty$, the calculation
would have proceeded as before with one difference. The fluctuations in
this case would originate overwhelmingly from the transverse modes,
since the number of distinct longitudinal modes is limited to the
number of vortices. However the completeness relation still holds, and
the condensed state may be taken as spatially homogeneous, thereby
enabling a calculation in the same spirit as that above. 

\section{Conclusions}

The main point of this article is that special care must be taken when
expanding the symmetry group of systems with spatially varying structures.
We have concentrated on one class of such systems, namely type-II 
superconductors in an external magnetic field. As mentioned in the
Introduction, there are applications of such models to heavy-Fermion
superconductors, and also to rotating superfluid He$_{3}$.
We have found a number of interesting results
connected with $N$-component superconductors, whose free energy
functional maintains a $U(N)$ symmetry. In general we have seen
that these systems adopt low-temperature configurations in 
which many OP components contribute by condensing into periodic 
structures -- leading
to a very rich structure for the overall condensate (and hence the
magnetic flux). 

We have examined the mean-field theory for the case $N=2$ in some detail.
The two OP components were found to condense into centered rectangular
structures, with an opening angle of $15^{\rm o}$. The two structures
are shifted relative to one another in such a way that the overall condensate
$|\phi_{1}|^{2}+|\phi_{2}|^{2}$ has a surprisingly rich structure, as shown
in fig.3. The generalized Abrikosov ratio for this configuration is 
$\beta_{g}(2) \simeq 1.0062$,
almost saturating the lower bound of unity. Although systems with
higher values of $N$ will adopt ever more complex structures, we were
able to show in section IV 
that the mean-field theory in the limit $N \rightarrow \infty$ 
has a simplifying feature, due to the completeness relation of the periodic
Eilenberger functions. Each OP component adopts an Eilenberger function,
but the overall condensate has no traces of the periodicity, and is 
spatially a constant. This structure saturates the lower bound of
$\beta_{g}(\infty)$.

In the remainder of 
section IV we examined the large-$N$ limit in more detail, by
considering a treatment which includes fluctuations self-consistently.
This calculation has been attempted several times in the 
past\cite{ba,lr,c1,c2}, but the previous authors have always assumed that
the system allows only one OP component to condense, which we have seen is
generically false. Our main finding is that for $d>4$, 
fluctuations do not disturb
the main characteristics found in mean-field theory -- namely a continuous
transition into a spatially homogeneous condensate, composed
of infinitely many OP components having condensed into Eilenberger functions.
[For $d<4$, the mixed phase is destroyed entirely.]
The system is found to have the critical properties
of the $O(2N)$ model of ferromagnetism\cite{ma,amit}, 
but in two fewer dimensions. Thus, 
exponents maintain their mean-field values above $d=6$, but take the 
values $\beta=1/2$ and $\nu = 1/(d-4)$ for $4<d<6$. The solution
we have found does not rely upon making the LLL approximation, or
upon neglecting gauge field fluctuations.

In this paper we have found that within mean-field theory, and also
in the large-$N$ limit, the $U(N)$ model undergoes a continuous
transition from the normal to the mixed phase. 
This is in contradiction to the results of some past works\cite{bnt,ba,c1},
the latter two of which contain errors of principle.
However, one may find a precedent for a continuous transition in
our recent study\cite{mn}, in which a functional renormalization
group (FRG) approach was applied to the $U(N)$ model via an expansion in
$\epsilon = 6-d$. In fact, the FRG study also predicted a mapping from
the $U(N)$ model to the $O(2N)$ model of ferromagnetism in two fewer
dimensions, for $N \ge 2$. It would be interesting to probe this relationship
further by extending the
present self-consistent analysis of the $U(N)$ model to finite $N$.
One of the more sophisticated means of achieving this would be via
the use of the parquet approximation\cite{ym}, which includes corrections
far beyond those of $O(1/N)$.

Finally we would like to draw the reader's attention to the fact that in
the exactly solvable large-$N$ limit, the mechanism for the transition
is the growing (phase) coherence in the direction of the applied field,
as the temperature is lowered. A theory of the ``melting'' transition
seen in high-temperature superconductors ($N=1$) has been recently
given by one of us\cite{moore}, based on the idea that the apparent
melting is just a consequence of crossover effects, when this phase
coherence length scale (which is very rapidly growing in three dimensions),
becomes comparable to the dimensions of the system.

\vspace{5mm}

\noindent
TJN acknowledges financial support from the Engineering and 
Physical Sciences Research Council.


\begin{references}

\bibitem[*]{na} address from 1st August 1997: Department of Physics,
Virginia Tech, Blacksburg, VA 24061, USA.

\bibitem{ma} S-K. Ma, Phys. Rev. Lett, {\bf 29}, 1311 (1972).

\bibitem{sig} M. Gell-Man and M. L\'evy, Nuovo Cim. {\bf 16}, 705 (1960).

\bibitem{signl} S. Coleman, R. Jackiw and H. D. Politzer, Phys. Rev. D,
{\bf 10}, 2491 (1974). 

\bibitem{wei} C. Y. Mou and P. Weichman, Phys. Rev. Lett. {\bf 70},
1101 (1993).

\bibitem{po} G. F. Mazenko and O. T. Valls, Phys. Rev. Lett. {\bf 51}, 2044
(1983); A. J. Bray, Adv. Phys. {\bf 43}, 357 (1994).

\bibitem{ir} J. P. Doherty et al, Phys. Rev. Lett. {\bf 72}, 2041 (1994).

\bibitem{ab} A. A. Abrikosov, Zh. Eskp. Teor. Fiz. {\bf 32}, 1442
(1957) [Sov. Phys. - JETP {\bf 5} 1174 (1957)].

\bibitem{hlm} B. I. Halperin, T. C. Lubensky and S-K. Ma, Phys. Rev. Lett.
{\bf 32}, 292 (1974). 

\bibitem{bnt} E. Br\'ezin, D. R. Nelson and A. Thiaville, Phys.\ Rev.\ B
  {\bf 31}, 7124 (1985).

\bibitem{ba} I. Affleck and E. Br\'ezin, Nucl. Phys. B {\bf 257},
  451 (1985).

\bibitem{lr} L. Radzihovsky, Phys. Rev. Lett. {\bf 74}, 4722 (1995).

\bibitem{c1} I. F. Herbut and Z. Te\v{s}anovi\'c, Phys. Rev. Lett. 
 {\bf 76}, 4450 (1996) (Comment).

\bibitem{c2} L. Radzihovsky, Phys. Rev. Lett. {\bf 76}, 4451 (1996) (Reply).

\bibitem{mn} M. A. Moore and T. J. Newman, Phys. Rev. Lett. {\bf 75}, 533
(1995); T. J. Newman and M. A. Moore, Phys. Rev. B {\bf 54}, 6661 (1996).

\bibitem{rev} M. Sigrist and K. Ueda, Rev. Mod. Phys. {\bf 63}, 239 (1991);
J. A. Sauls, Adv. Phys. {\bf 43}, 113 (1994) and references therein.

\bibitem{he3} D. R. T. Jones, A. Love and M. A. Moore, J. Phys. C {\bf 9}, 
743 (1976); D. Bailin, A. Love and M. A. Moore, J. Phys. C {\bf 10}, 1159 
(1977).

\bibitem{fh} A. L. Fetter and P. C. Hohenberg, in {\it Superconductivity},
ed. R. D. Parks (Dekker, New York, 1969), Vol. 2.

\bibitem{llqm} L. D. Landau and E. M. Lifshitz, {\it Quantum Mechanics} 
(Pergamon, London, 1958).

\bibitem{ge} G. Eilenberger, Phys. Rev. {\bf 164}, 628 (1967).

\bibitem{mf2op} A. Garg and D-C. Chen, Phys. Rev. B {\bf 49}, 479 (1994).

\bibitem{cw} S. Coleman and E. Weinberg, Phys. Rev. D {\bf 7}, 1888 (1973).

\bibitem{amit} D. J. Amit, {\it Field Theory, the Renormalization Group
and Critical Phenomena} 2nd edition (McGraw-Hill, New York, 1988).

\bibitem{mn1} M. A. Moore and T. J. Newman, unpublished.

\bibitem{ym} J. Yeo and M. A. Moore, Phys. Rev. Lett. {\bf 76}, 1142 (1996);
Phys. Rev. B {\bf 54}, 4218 (1996).

\bibitem{moore} M. A. Moore, Phys. Rev. B {\bf 55}, 14136 (1997). 


\end{references}
\end{document}